\documentclass[sn-mathphys,Numbered]{sn-jnl}

\usepackage{graphicx}%
\usepackage{multirow}%
\usepackage{amsmath,amssymb,amsfonts}%
\usepackage{amsthm}%
\usepackage{mathrsfs}%
\usepackage[title]{appendix}%
\usepackage{xcolor}%
\usepackage{textcomp}%
\usepackage{manyfoot}%
\usepackage{booktabs}%
\usepackage{algorithm}%
\usepackage{algorithmicx}%
\usepackage{algpseudocode}%
\usepackage{listings}%

\theoremstyle{thmstyleone}%
\theoremstyle{thmstyletwo}%

\theoremstyle{thmstylethree}%

\usepackage{ulem}

\begin{document}

\title[Chaotic dynamics of pulsating spheres orbiting black holes]{Chaotic dynamics of pulsating spheres orbiting black holes}


\author*[1,2]{\fnm{Fernanda} \sur{de F. Rodrigues}}\email{fernanda@ihep.ac.cn}

\author[1,3]{\fnm{Ricardo} \sur{A. Mosna}}\email{mosna@unicamp.br}

\author[4]{\fnm{Ronaldo} \sur{S. S. Vieira}}\email{ronaldo.vieira@ufabc.edu.br}

\affil[1]{\orgdiv{Instituto de F{\'i}sica Gleb Wataghin}, \orgname{ Universidade Estadual de Campinas}, \orgaddress{\street{S{\'e}rgio Buarque de Holanda}, \city{Campinas}, \postcode{13083-859}, \state{S\~ao Paulo}, \country{Brazil}}}

\affil[2]{\orgdiv{Institute of High Energy Physics}, \orgaddress{\street{19B Yuqhan Road}, \city{Beijing}, \postcode{100049}, \country{China}}}

\affil[3]{\orgdiv{Departamento de Matem\'atica Aplicada}, \orgname{Universidade Estadual de Campinas}, \orgaddress{\street{S{\'e}rgio Buarque de Holanda}, \city{Campinas}, \postcode{13083-859}, \state{S\~ao Paulo}, \country{Brazil}}}

\affil[4]{\orgdiv{Centro de Ci\^encias Naturais e Humanas}, \orgname{Universidade Federal do ABC}, \orgaddress{\street{Avenida dos Estados}, \city{Santo Andr\'e}, \postcode{09210-580}, \state{S\~ao Paulo}, \country{Brazil}}}


\abstract{We study the chaotic dynamics of spinless extended bodies in a wide class of spherically symmetric spacetimes, which encompasses black-hole scenarios in many modified theories of gravity. We show that a spherically symmetric pulsating ball may have chaotic motion in this class of spacetimes. The cases of the Reissner-Nordstr{\"o}m and Ay{\'o}n-Beato-Garc{\'i}a black holes are analyzed in detail. The equations of motion for the extended bodies are obtained according to Dixon's formalism, up to quadrupole order. Then, we use Melnikov's method to show the presence of homoclinic intersections, which imply chaotic behavior, as a consequence of our assumption that the test body has an oscillating radius.}




\maketitle

\section{Introduction}\label{Introduction}

The study of chaotic trajectories in a physical system whose integrability is broken by means of a perturbation in its background is a well-established and classic subject of nonlinear dynamics. The investigation of regular and chaotic dynamics in astrophysics has a long tradition \cite{barrow1997poincare, binneytremaineGD, contopoulosOCDA2002}, with applications ranging from planetary \cite{barrow1997poincare, MurrayDermott2000, lecarEtal2001ARAA, ferrazmeloMichtchenkoEtal2005LNP, vieiraRamosCaro2023NewA} to galactic \cite{binneytremaineGD, contopoulosOCDA2002, henonheiles1964AJ, contopoulos2003galaxies, grosbol2003LNP, grosbol2002SSRv, hunter2003LNP, hunter2005NYASA, quillen2003AJ, ramoscaroLopezsuspesGonzalez2008MNRAS, zotos2011CSF, pichardoEtal2003ApJ, michtchenkoVieiraBarrosLepine2017AA, lepineVieiraEtal2017ApJ, michtchenkoVieiraEtal2018ApJL, contopoulos1960ZA, contopoulos1963AJ, contopoulos1967rta2, dezeeuw1985MNRAS, dezeeuw1988NYASA, binneyMcmillan2011MNRAS, binney2010MNRAS, binney2012MNRASdynamical, binney2012MNRASactions, binneySanders2014IAUS, bienaymeRobinFamaey2015AA, vieiraRamosCaro2016CeMDA, vieiraRamoscaro2014ApJ, vieiraRamoscaro2019MNRAS, vieiraRamoscaroSaa2016PRD} and cosmological \cite{barrow1982PhysRep, chernoffBarrowPhysRevLett1983, contopoulosEtal1999, motterLetelier2001PLA} scales. 

Regarding orbital dynamics in general relativity, chaotic geodesic motion appeared in many different contexts throughout the years, being analyzed by diverse methods, e.g. \cite{hobill1994deterministic,bombelli1992CQGra, letelierVieira1997CQGra, saaVenegeroles1999PhLA, lukesgerakopoulos2014PhRvD, seyrichEtal2012PhRvD, saltasEta2024arXiv240212359S}. The dynamics of test particles in relativistic systems has been extensively studied, with chaos identified in numerous scenarios. We mention below some of its applications to relativistic systems, though this is not intended to be a review of the literature. Chaos was found, for instance, in ``black hole + gravitational wave'' spacetimes \cite{bombelli1992CQGra, letelierVieira1997CQGra}, spacetimes representing the exterior field of quadrupole-deformed \cite{gueronLetelier2001PRE, gueronLetelier2002PRE} and other nonspherical \cite{lukesgerakopoulos2012PhRvD} central bodies, and in axisymmetric gravitating disk systems with or without a central black hole \cite{saaVenegeroles1999PhLA, wuZhang2006ApJ, semerakSukova2010MNRAS, semerakSukova2012MNRAS, sukovaSemerak2013MNRAS, witzanySemerakSukova2015MNRAS, polcarSukovaSemerak2019ApJ}. 

A recent advancement in the field regards the chaotic dynamics of inspiralling stellar-mass compact objects into supermassive black holes (extreme-mass-ratio binary inspirals). Assuming that the supermassive black hole is not described by the Kerr geometry \cite{apostolatos2009PhRvL, contopoulosEtal2011IJBC, mukherjeeEtal2023PhRvD, destounisKokkotas2023GReGr}, the resulting motion is generally non-integrable, leading to chaotic behavior in the inspiraling object. This chaotic nature would imprint signatures in the generated gravitational-wave signals \cite{lukesgerakopoulosEtal2010PhRvD, zelenkaEtal2020PhRvD, destounisEtal2020PhRvD, destounisEtal2021PhRvL, destounisEtal2021PhRvD, destounisEtal2023GReGr, destounisEtal2023PhRvD, eleniEtal2024arXiv240802004E}, potentially detectable by future gravitational-wave observatories such as the Laser Inteferometer Space Antenna (LISA) \cite{lisa2023LRR}. 

Furthermore, the orbital dynamics of classical spinning particles was also shown to be generally chaotic, even in Schwarzschild spacetime \cite{suzukiMaeda1997PRD, zelenkaEtal2020PhRvD}. Since the particle's spin is a dynamic quantity that couples with its four-momentum in the equations of motion \cite{papapetrou1951RSPSA},  the resulting dynamical system possesses a higher dimensionality compared to the geodesic case. This increase in dimensionality explains the general lack of regularity when the spin around the particle's center of mass is considered.

We have recently explored a different route to chaos, in the dynamics of extended bodies, which arises due to finite-size corrections to the otherwise integrable motion of a test particle. This was done both in Newtonian mechanics~\cite{bolinha} and in general relativity, in the context of Schwarzschild spacetime~\cite{papersch}. In the latter case, we showed that the trajectory of a nearly spherical body may become chaotic when its shape oscillates between a prolate and an oblate spheroid.

In this paper we show that, for spacetimes with metric
\begin{equation}\label{metric}
ds^2=-f(r)\, dt^2+f(r)^{-1}dr^2+ r^2(d\theta^2+ \sin^2 \theta\, d\phi^2),
\end{equation}
this effect may arise even if the extended body is spherically symmetric at all times, {\it e.g.}, if the body is a pulsating sphere. This form of the metric describes many black-hole solutions in modified theories of gravity (see, for instance, \cite{ayonbeatoGarcia1998PRL, ansoldi2008arXiv, kehagiasSfetsos2009PhLB, stashkoZhadanov2018GRG, vagnozziEtal2023CQG} and references therein), since it automatically guarantees the necessary condition \cite{carter2009GRG, carter2010GRG} that the event horizon $g^{rr}=0$ is also a Killing horizon for $\partial_t$ (having $g_ {tt}=0$). 
We find that a pulsating spherical ball follows a geodesic in vacuum (with or without cosmological constant), but that homoclinic chaos arises near unstable circular orbits in more general spherically symmetric spacetimes. Regarding the behavior of matter, we only require that it satisfies the usual conservation laws and that test particles follow geodesics.
As examples of this phenomenon we analyze the cases of the Reissner-Nordstr{\"o}m and Ay{\'o}n-Beato-Garc{\'i}a black-hole spacetimes.

This work is organized as follows. In Section \ref{Dixon}, after reviewing some aspects of Dixon's formalism for extended bodies up to quadrupole order, we set up the equations of motion of oscillating spherically symmetric bodies in spacetimes with metric (\ref{metric}).
In Section \ref{secmel} we analyze the Melnikov function of the system (associated with the corresponding homoclinic orbits), which allows us to determine conditions for chaotic motion around Reissner-Nordstr{\"o}m and Ay{\'o}n-Beato-Garc{\'i}a black holes. We present our conclusions in Section \ref{sec:conclusion}.

\section{Dixon's formalism for spherically symmetric bodies}
\label{Dixon}

Consider an extended test body defined by an energy-momentum tensor $T_{\mu\nu}$ in a spacetime with metric $g_{\mu\nu}$. Dixon's formalism replaces the study of the conservation equations
\begin{equation}\label{divT=0}
\nabla_{\mu}T^{\mu\nu}=0
\end{equation}
by a set of ordinary differential equations  for the evolution of the linear momentum $p_{\mu}(\tau)$ and spin $S^{\mu\nu}(\tau)$ of the body:
\begin{eqnarray}
\label{eq:p&S}
\frac{Dp^{\mu}}{d\tau} &=& -\frac{1}{2}R^{\mu}_{\phantom{\mu} \nu \alpha \beta}v^{\nu}S^{\alpha \beta}+F^{\mu}, \label{eq:p} \\
\frac{DS^{\mu \nu}}{d\tau} &=& 2p^{[\mu}v^{\nu]}+N^{\mu \nu}, \label{eq:S}
\end{eqnarray}
where $R^{\mu}_{\phantom{\mu} \nu \alpha \beta}$ is the Riemann tensor of $g_{\mu\nu}$~\cite{DixonI,DixonII,DixonIII}. Here $v^\mu=dz^\mu/d\tau$, with  $z^\mu(\tau)$ the body's center-of-mass position, which is implicitly defined by the condition
\begin{equation}
S^{\mu \nu}(\tau) p_{\nu}(\tau)=0.
\end{equation}
We take the evolution parameter $\tau$ as the proper time along the center-of-mass trajectory.\footnote{We remark that $p^\mu$ and $v^\mu$ are, in general, not parallel.}

The force $F^{\mu}$ and torque $N^{\mu \nu}$ are obtainable from a multipole expansion of the energy-momentum conservation relation (\ref{divT=0}) for the body. Their expressions up to quadrupole order are given by
\begin{eqnarray}
\label{eq:F&N}
F^{\mu} &=&- \frac{1}{6}J^{\alpha \beta \gamma \delta}\nabla^{\mu}R_{\alpha \beta \gamma \delta}, \label{eq:F} \\
N^{\mu \nu} &=&\frac{4}{3}J^{\alpha \beta \gamma[\mu}R^{\nu]}_{\phantom{ai} \gamma \alpha \beta}, \label{eq:N}
\end{eqnarray}
where $J^{\alpha \beta \gamma \delta}$ is the quadrupole moment of the body~\cite{DixonI,DixonII,DixonIII}, which can be chosen to have the same algebraic properties as the Riemann tensor,
\begin{equation}
\label{Jsymmetries}
J^{\alpha \beta \gamma \delta}=J^{[\alpha \beta] \gamma \delta}=J^{\alpha \beta [\gamma \delta]}, \quad \quad J^{[\alpha \beta \gamma] \delta}=0.
\end{equation}

Perhaps the most appealing feature of Dixon's formalism is its judicious choice of the spacetime foliation according to which the multipoles are expanded.
It allows $J^{\alpha \beta \gamma \delta}$ to be freely prescribed as a function of $\tau$, with no other restrictions apart from the usual energy conditions.
The same is true for the higher order (octupole, etc) multipoles. At the quadrupole level, all the information about the internal structure of the body is contained in $J^{\alpha \beta \gamma \delta}$, which may be thought of as being freely prescribed by internal mechanisms of the body.%

If $\{e_a\}$ is a frame comoving with the body, it follows from the hypothesis that the body is spherically symmetric and from equations (\ref{Jsymmetries}) that only six components of the quadrupole tensor may be nonzero:
\begin{eqnarray}
j_m :=& J_{0101}=J_{0202}=J_{0303}, \label{eqjm}\\
j_s : =&  J_{2323}=J_{1313}=J_{1212}. \label{eqjs}
\end{eqnarray}
The components $J_{0i0i}$ and $J_{ijij}$, $i,j=1,2,3$, are, respectively, the mass and stress quadrupole moments of the body~\cite{Ehlers1977DynamicsOE}.

We are interested in the motion of test bodies in spacetimes described by the metric (\ref{metric}) when the body has no spin and  hence, by symmetry, has its motion confined to a plane, which we take as $\theta=\pi/2$.
Also by symmetry, a spherical test body which starts with zero spin will remain spinless at all times and we are only left to deal with the translational degrees of freedom of the body's center of mass.
The relevant equations of motion then reduce to equation (\ref{eq:p}) which now simplifies to
\begin{equation}
\label{eq:psimplificada}
\frac{Dp_\mu}{d\tau}=F_\mu,
\end{equation}
for $\mu=t$, $r$, $\phi$.  As a consistency check, we can show that these equations imply that equation (\ref{eq:S}) is automatically satisfied for $S^{\mu\nu}=0$ in the present case of a spherically symmetric body in a spherically symmetric spacetime of the form (\ref{metric}).
Calculating the force (\ref{eq:F}) in the frame $\{e_a\}$ and transforming it back to the coordinate basis (in the same manner as done in reference \cite{papersch}), we obtain the following nonzero components for $F_\mu$:
\begin{eqnarray}\label{eq:forcavacuo}
F_r&=&-\frac{\left(j_{s}C+j_{m}f(r)\,p_{\phi}^2\right)\,\beta+\left(j_{s}f(r)\,p_{\phi}^2+j_{m}C\right)\,\alpha}{3r^3\left[f(r)\left( p_{\phi}^2+f(r)\,p_r^2\,r^2\right)-p_t^2\,r^2\right]},\\
F_{\phi}&=&-\frac{(j_{s}+j_{m})\,p_{\phi}f(r)^2p_r\,\beta}{3r\left[f(r)\left(p_{\phi}^2+f(r)\,p_r^2\,r^2\right)-p_t^2\,r^2\right]},
\end{eqnarray}
where \mbox{$C=r^2[p_t^2+p_r^2f(r)^2]$}, \mbox{$\beta=2[2-2f(r)+r^2f''(r)]$} and $\alpha=r[2f'(r)-r(2f''(r)+rf^{(3)}(r))]$.

It is interesting to apply this result to the case when the spacetime is that of a spherically symmetric black hole in vacuum, possibly along with a cosmological constant, i.e., the Schwarzschild-(A)dS black hole,
\begin{equation}
\label{metric SAdS}
f(r)=1-\frac{2M}{r}\pm\frac{r^2}{\ell^2}.
\end{equation}
In this case, both $\alpha$ and $\beta$ are identically zero and therefore $F_\mu$ is identically zero. This shows (at quadrupolar order) that the center-of-mass trajectory of a spherically symmetric ball, pulsating or not, is always a geodesic on these spacetimes.

A less impressive simplification in the form of $F_\mu$ also happens for the Reissner-Nordstr\"om metric, corresponding to equation (\ref{metric}) with
\begin{equation}
\label{RN metric}
f(r)=1-\frac{2M}{r}+\frac{Q^2}{r^2},
\end{equation}
where $M$ and $Q$ are the mass and charge of the black hole. In this case, $\alpha=\beta=8 Q^2/r^2$, so that the force $F_\mu$ depends on the quadrupole moment only through $j_m+j_s$.

\subsection{Canonical equations of motion at quadrupolar order}
\label{eqofmot}

It follows from Dixon's formalism that each Killing vector $\xi$ of the spacetime gives rise to a conserved quantity~\cite{DixonI}
\begin{equation}
\label{eq:cons_qtity}
\mathscr{P}_{\xi}=p_{\mu}\xi^{\mu}+ \frac{1}{2}S^{\mu \nu}\nabla_{\mu}\xi_{\nu}.
\end{equation}
In our case, the spacetime is described by the metric (\ref{metric}) for an arbitrary, positive $f(r)$. Therefore  $\xi=\partial_t$ and $\xi=\partial_\phi$ are Killing fields and thus $p_t$ and $p_\phi$ are constants of motion.

Therefore, we can write equations (\ref{eq:psimplificada}) as
\begin{eqnarray}
\label{eq:U=something}
\frac{dt}{d\tau} &=& h_1(r,p_r,\dot{p}_r,\tau), \label{eq:Ut=something}\\
\frac{dr}{d\tau} &=& h_2(r,p_r,\dot{p}_r,\tau), \label{eq:Ur=something}\\
\frac{d\phi}{d\tau} &=& h_3(r,p_r,\dot{p}_r,\tau), \label{eq:Uphi=something}
\end{eqnarray}
where $p_t$ and $p_\phi$ only appear as fixed parameters~\cite{papersch}. These equations depend explicitly on $\tau$ only via the quadrupole moment $J^{\alpha \beta \gamma \delta}(\tau)$. Also note that, by symmetry, $h_1$, $h_2$ and $h_3$ are not functions of $t$ and $\phi$ (this can of course be explicitly checked). We then proceed as in \cite{papersch} and substitute equations (\ref{eq:U=something})--(\ref{eq:Uphi=something}) into $v_\mu v^\mu=-1$ to get an equation of the form $\dot{p}_r=h_4(r,p_r,\tau)$. After substituting this back into equation (\ref{eq:Ur=something}) we get an equation of the form $\dot{r}=h_5(r,p_r,\tau)$. As a result, we obtain the following non-autonomous dynamical system:
\begin{eqnarray}
\label{eq:sistdyn}
\dot{p}_r &= h_4(r,p_r,\tau), \\
\dot{r} &= h_5(r,p_r,\tau),
\end{eqnarray}
describing the radial translational motion of the extended body (the azimuthal motion is then given by conservation of $p_\phi$).

After a lengthy calculation, this leads, up to quadrupolar order, to
\begin{eqnarray}
\label{eq:sistdyn2}
\frac{dr}{d\tau}      =& f_1(r,p_r) + j_m(\tau) g_{1m}(r,p_r) + j_s(\tau)\,  g_{1s}(r,p_r), \label{eq:sistdyn2a} \\
\frac{dp_r}{d\tau}  =& f_2(r,p_r) + j_m(\tau)\,  g_{2m}(r,p_r)+j_s(\tau)\,  g_{2s}(r,p_r). \label{eq:sistdyn2b}
\end{eqnarray}
The explicit expressions for each of these functions are shown in Appendix \ref{appendix}, with $E=-p_t$ and $L=p_\phi$.

A direct calculation then shows that
\begin{eqnarray}
\label{eq:constraintsfg}
\quad\frac{\partial f_1}{\partial r}+\frac{\partial f_2}{\partial p_r}\! &= 0,
\label{eq:constraintsfg a}\\
\frac{\partial g_{1m}}{\partial r}+\frac{\partial g_{2m}}{\partial p_r} &= 0, \label{eq:constraintsfg b}\\
\,\,\frac{\partial g_{1s}}{\partial r}+\frac{\partial g_{2s}}{\partial p_r}\! &= 0, \label{eq:constraintsfg c}
\end{eqnarray}
which means that this dynamical system is Hamiltonian with $r$ and $p_r$ being canonically conjugate variables.

\subsection{Test-particle orbits}
\label{homorb}

In the limit when the body is a point particle, its equations of motion are the geodesic equations. This may be obtained from equations (\ref{eq:sistdyn2}) as follows. We write
\begin{equation}
m^2=-p_\mu p^\mu,
\end{equation}
where $m$ is the mass of the body (which is in general not constant when the body is extended~\cite{DixonI,DixonII,DixonIII}).
The momentum $p_r$ can then be expressed in terms of $r$ by solving the above equation for $p_r$,
\begin{equation}
p_r^2=\frac{1}{f(r)} \left(
\frac{E^2}{f(r)}-\frac{L^2}{r^2}-m^2
\right),
\end{equation}
and noting that, for a point particle, $m$ is constant, $m=m_0$. Substituting this result in equation (\ref{eq:sistdyn2a}) with $j_m=j_s=0$, we obtain
\begin{equation}
\label{eq:point_particle}
\frac{\dot{r}^2}{2} + \frac{f(r)}{2}\left( 1+\frac{l^2}{r^2}\right) = \frac{e^2}{2},
\end{equation}
where $e=E/m_0$ and $l=L/m_0$ (the specific energy and angular momentum) are also constants.
This is the usual conservation equation for an equatorial orbit of a point particle on a spacetime with metric (\ref{metric}). It may be written in terms of the ``effective potential''
\begin{equation}
\label{eq:Veff}
V_{\rm eff}=\frac{f(r)}{2}\left( 1+\frac{l^2}{r^2} \right),
\end{equation}
whose local minima and maxima give rise to stable and unstable circular orbits, respectively (for each fixed value of $l$).

We note that, by Rayleigh's stability criterion for circular orbits, there will always be an instability region for these orbits in the outer vicinity of any unstable light ring \cite{vieiraWlodekMarek2017PRD}. If the spacetime is asymptotically flat, then the behavior near infinity is Keplerian and therefore there will be a region of stability for circular orbits at larger radii, implying the existence of a marginally stable circular orbit outside the photon ring (these results can also be obtained by explicit calculations of the perturbative motion around timelike circular geodesics \cite{delgadoEtal2022PRD}). In this way, if these unstable circular orbits have $e<1$, thus lying in a bounded energy level set, the dynamical system will contain, for each $l$ above a certain threshold, a homoclinic orbit corresponding to this unstable fixed point. This is the case of Schwarzschild, Reissner-Nordstr\"om and Ay{\'o}n-Beato-Garc{\'i}a spacetimes, for instance.

\subsection{Extended body and quadrupolar oscillations}
\label{chquad}

We now go back to the case of an extended body. We consider a time-dependent profile for $j_m$, oscillating with a given frequency $\Omega$,
\begin{equation}\label{eq:js}
j_m(\tau)=q_0\left[1+\sin(\Omega\tau)\right],
\end{equation}
with $j_m$ always nonnegative, since it is a mass quadrupole. The extended body is then a spherical ball with an oscillating  radius. For simplicity, we take $j_s(\tau)=0$. This is just a simplifying assumption which corresponds to the case when the mass quadrupoles are dominant over the stress quadrupoles. However, we see from equations (\ref{eq:sistdyn2}) that our formalism does allow nontrivial configurations for the stress quadrupole as well.

We show in the next Section that this simple kind of perturbation in the extended-body structure already leads to chaotic motion when the unperturbed orbit (the point-particle limit) is chosen as a homoclinic orbit.

\section{Homoclinic Chaos}
\label{secmel}

In the presence of a homoclinic orbit of the unperturbed, autonomous two-dimensional dynamical system (\ref{eq:sistdyn2})--(\ref{eq:sistdyn2b}), time-periodic perturbations to the dynamics may break its integrability. One way to detect this phenomenon is by means of Melnikov's method \cite{holmes1990PhysRep, guckenheimerHolmes2013, lichtenbergLieberman1992}, which provides an algorithmic way to search for homoclinic intersections between the stable and unstable manifolds of the perturbed fixed point in the Poincar\'e stroboscopic map given by $\tau_n = 2 n \pi/\Omega$, with $n$ being an integer, where $\Omega$ is the fundamental frequency of the perturbation.

We write $\dot r = f_1(r,p_r) + \epsilon \lambda_1(r,p_r, \tau)$, $\dot p_r = f_2(r,p_r) + \epsilon \lambda_2(r,p_r, \tau)$ with $\epsilon$ small and $\lambda_i$ periodic in $\tau$ with frequency $\Omega$. If $f_1$ and $f_2$ satisfy equation (\ref{eq:constraintsfg a}), then Melnikov's integral
\begin{equation}\label{eq:melnikovIntegral}
M(\tau_0)=\int_{-\infty}^{\infty}(f_1\lambda_2-f_2\lambda_1)(r(\tau),p_r(\tau),\tau+\tau_0)\,d\tau
\end{equation}
is proportional to the first-order term in the transversal distance between these stable and unstable manifolds, evaluated with respect to the unperturbed homoclinic orbit at $\tau=\tau_0$. Here in equation (\ref{eq:melnikovIntegral}) the trajectory $(r(\tau), p_r(\tau))$ is evaluated along the unperturbed homoclinic orbit. Therefore, simple zeros of $M(\tau_0)$ represent transversal crossings of the manifolds, giving rise to a homoclinic tangle in phase space \cite{holmes1990PhysRep, lichtenbergLieberman1992} and to chaotic dynamics around this region.%
\footnote{As in the orbital analysis of changing-shape bodies in Schwarzschild spacetime \cite{papersch}, here chaos is transient due to the body's infall into the black hole. For more details, see \cite{papersch} and references therein.}
In our case, $\epsilon \lambda_1=j_m(\tau)g_{1m}$ and $\epsilon \lambda_2=j_m(\tau)g_{2m}$, with $j_m(\tau)$ given by equation (\ref{eq:js}) and $f_1$, $f_2$, $g_{1m}$, $g_{2m}$ by equations (\ref{eq:f1})--(\ref{eq:g2s}). Here, $q_0$ corresponds to the perturbation parameter $\epsilon$.

We now consider three examples of spherically symmetric black hole spacetimes and the specific form that the relations (\ref{eq:sistdyn2}) assume in each case.

\subsection{Schwarzschild-(A)dS spacetime}
\label{secads}

In this case, as anticipated before, the orbital motion of the extended body is unchanged, as compared to the point-particle case, for any form of $j_s$ and $j_m$. In terms of equations (\ref{eq:sistdyn2}), this happens because $g_{1m}$, $g_{1s}$, $g_{2m}$, and $g_{2s}$ are all identically zero when $f(r)$ is given by equation (\ref{metric SAdS}).

\subsection{Reissner-Nordstr\"om spacetime}
\label{secrn}

In the Reissner-Nordstr\"om metric, characterized by the function $f(r)$ given by equation (\ref{RN metric}), the horizons are located at $r=r_{h \pm}$ with $r_{h \pm}=M \pm (M^2-Q^2)^{1/2}$. As an immediate consequence, the analysis of the possible stable and unstable circular orbits depends on the charge-to-mass ratio, $Q/M$ \cite{puglieseQuevedoRuffini2011PRD, katkaVieiraEtal2015GRG}. Here, we focus on the subcritical regime, defined by $0 < Q < M$, which corresponds to a black hole spacetime. The case $Q=M$ --- the extremal black hole case --- is qualitatively similar for the problem we are interested in.

\begin{figure}
	\begin{center}
		\includegraphics[width=0.47\columnwidth]{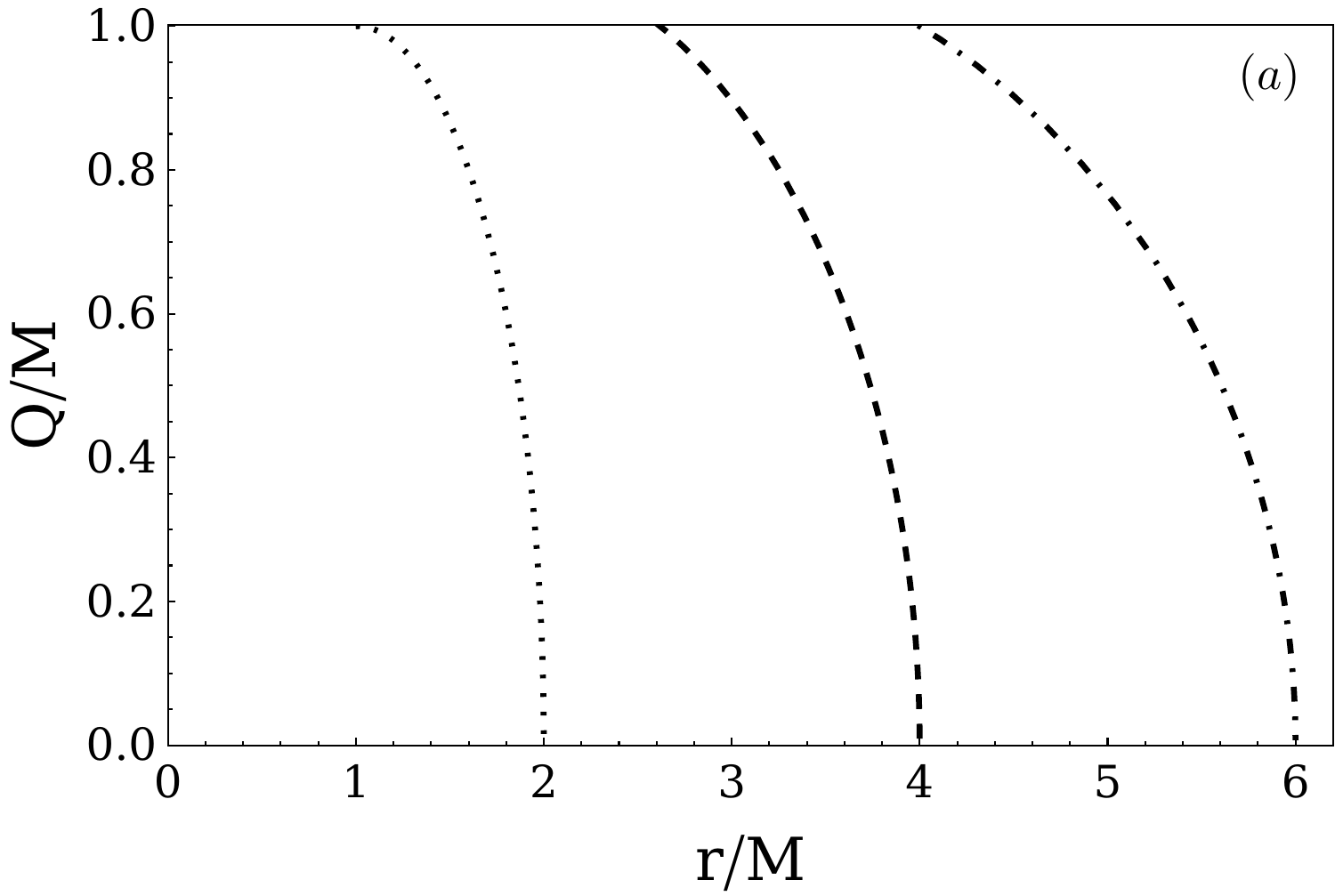}\qquad
		\includegraphics[width=0.47\columnwidth]{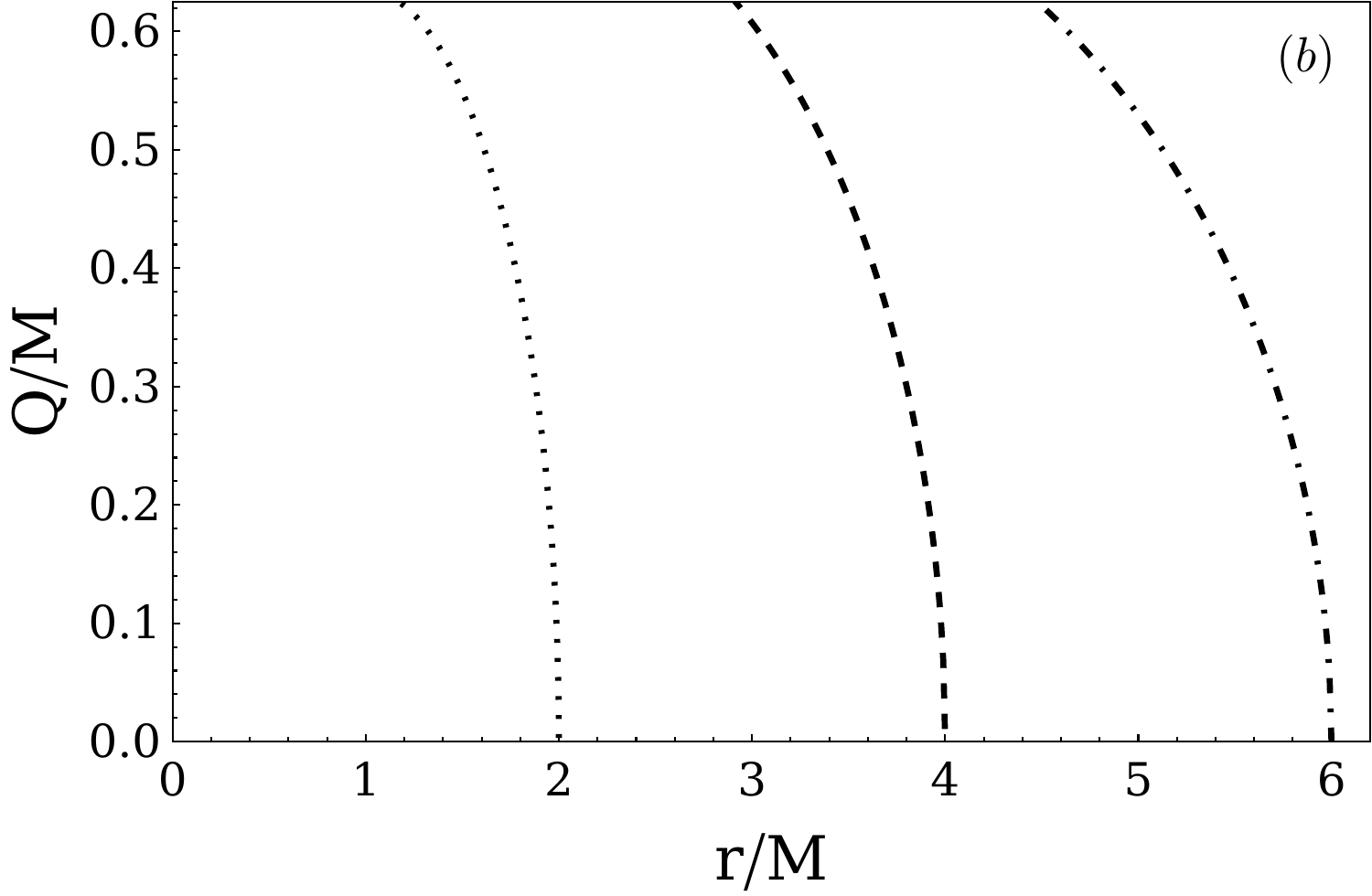}
		\caption{ (a) Plot of $r_{h +}/M$ (dotted line), $r_{boun}/M$ (dashed line) and $r_{isco}/M$ (dash-dotted line) for the
		 Reissner-Nordstr\"om metric with $0 < Q/M < 1$.
		(b) Plot of $r_{+}/M$ (dotted line), $r_{boun}/M$ (dashed line) and $r_{isco}/M$ (dash-dotted line) for the
		 Ay{\'o}n-Beato-Garc{\'i}a metric with  $0< Q/M < Q_c/M \approx 0.634$. 
		}
		\label{fig:orbitas}
	\end{center}
\end{figure}

Regarding test-particle motion, for each value of angular momentum $l$ above a certain threshold there is an unstable circular orbit at $r=r_{un}$, which has a homoclinic orbit associated with it if $e<1$. The allowed range for $r_{un}$ is given by $r_{boun}<r_{un}<r_{isco}$, where $r_{isco}$ is the marginally stable circular orbit radius and $r_{boun}$ is the marginally bound ($e=1$) circular orbit radius. The locations of $r_{boun}/M$ (dashed line), $r_{isco}/M$ (dash-dotted line), and $r_{h +}/M$ (dotted line) for different values of the charge-to-mass ratio are shown in figure~\ref{fig:orbitas}(a).

\begin{figure}
	\begin{center}
		\includegraphics[width=0.47\columnwidth]{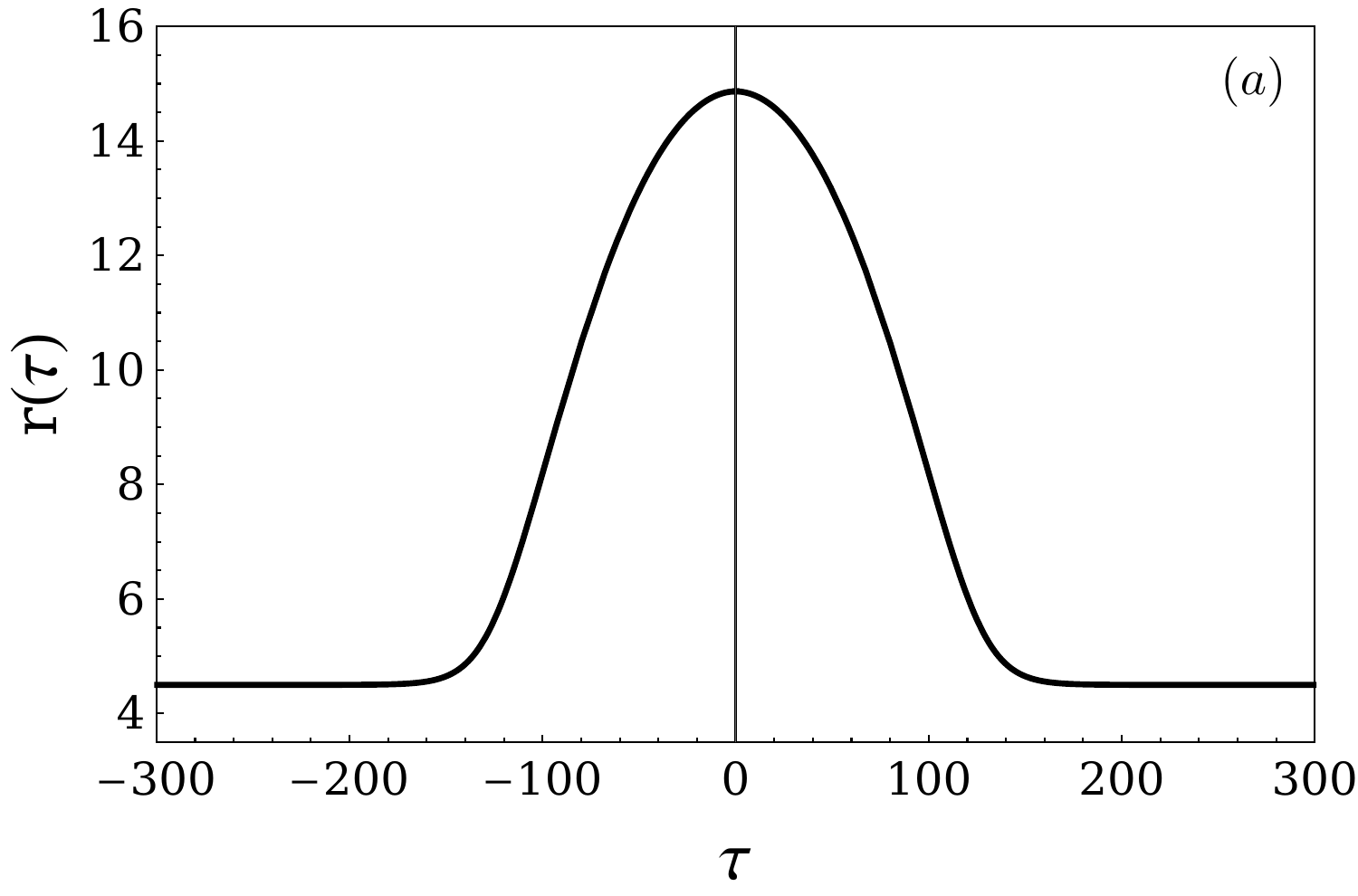}\qquad
		\includegraphics[width=0.47\columnwidth]{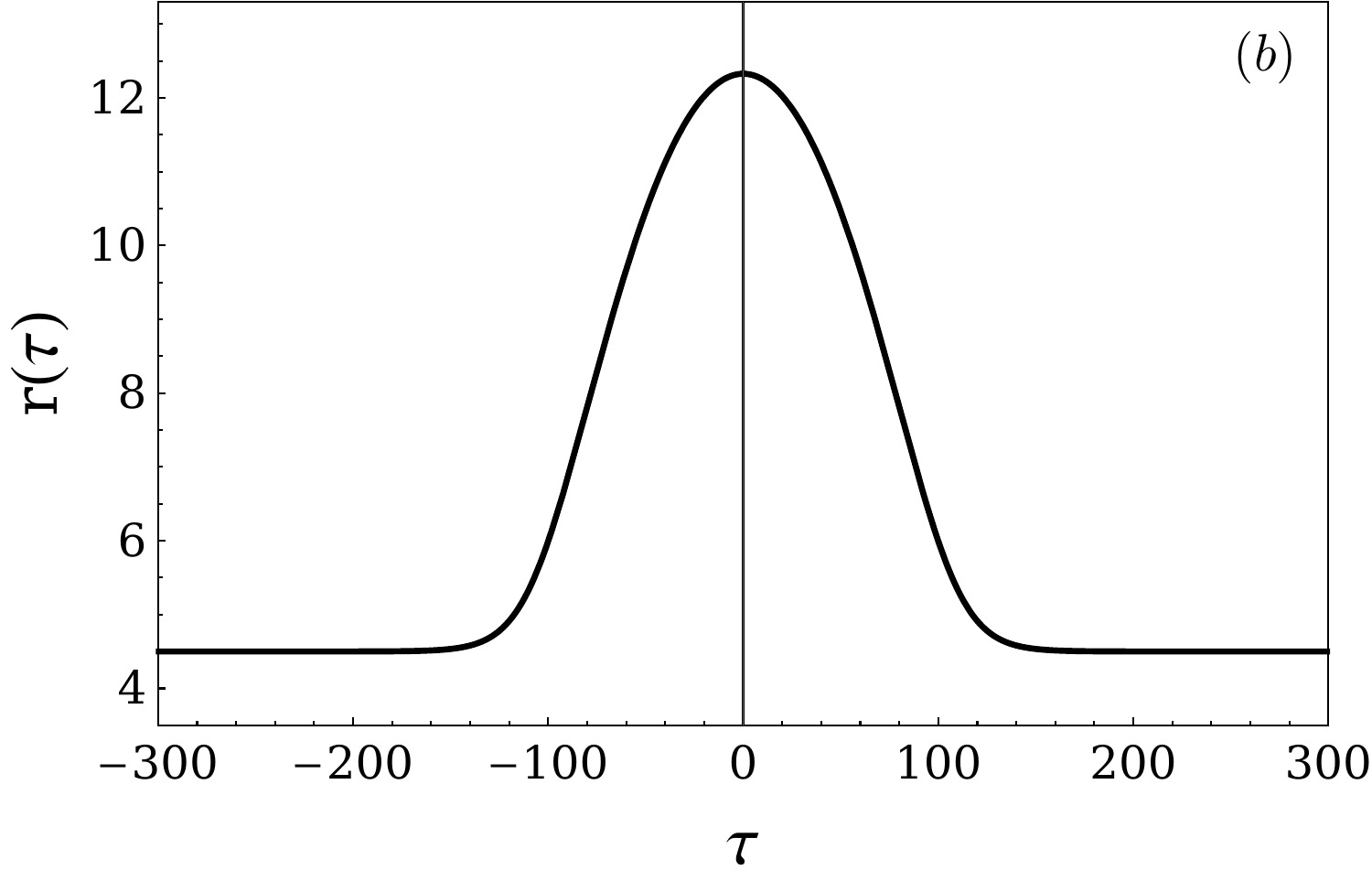}
		\caption{ Plot of $r(\tau)$, with $M=1$, $Q=0.3$ and $r_{un}=4.5$ for a homoclinic orbit in the
		(a) Reissner-Nordstr\"om metric and the
		(b) Ay{\'o}n-Beato-Garc{\'i}a metric.
		}
		\label{fig:homoclinica}
	\end{center}
\end{figure}

The energy $e$ and angular momentum $l$ of the unperturbed homoclinic orbit can be written as functions of $r_{un}$, the mass $M$, and charge $Q$ of the black hole. This allows the corresponding conservation equation (\ref{eq:point_particle}) to be written as follows
\begin{equation}
 \label{eq:homoclinica}
 \frac{dr}{d\tau}=\pm \frac{r-r_{un}}{r^2\,r_{un}}\sqrt{\frac{Q^4b^2+Mr_{un}\left[r\,r_{un}c-Q^2a^2\right]}{2Q^2+r_{un}(r_{un}-3M)}}\,,
\end{equation}
where $a=2r+r_{un}$, $b=r+r_{un}$ and $c=2Ma-r\,r_{un}$. The plot of $r$ versus $\tau$ is presented in figure~\ref{fig:homoclinica}(a). The constant of integration was chosen so that $\tau(r=r_m)=0$, with $r_m$ being the turning point
of the homoclinic orbit. Note that $r(\tau)$ tends asymptotically to $r_{un}$.

The solution to equation (\ref{eq:homoclinica}) corresponds to the unperturbed homoclinic orbit in phase space, which we expect to break in the quadrupolar approximation. To investigate this behavior we write equation (\ref{eq:melnikovIntegral}) with $f_1$, $f_2$, $g_{1m}$, and $g_{2m}$ given by equations (\ref{eq:f1}) -- (\ref{eq:g2s}), resulting in a Melnikov integral of the form
\begin{equation}
 \label{eq:melnikovrn}
 M(\tau_0)=2\cos{(\Omega \tau_0)}K(\Omega)
\end{equation}
where
\begin{equation}
\label{eq:integralrn}
 K(\Omega)=\int_{r_{un}}^{r_m}k(r)\sin{[\Omega \tau(r)]}\,dr.
\end{equation}
The function $k(r)$ is given by equation (\ref{eq:kappa}), and
the behavior of $K$ as a function of $\Omega$ can be seen in figure 3(a), with $M=1$, $Q=0.3$ and $r_{un}=4.5$. We see that $K$ has isolated zeros in the interval $0<\Omega <0.3$.
Therefore, for values of $\Omega$ for which $K \ne 0$, the Melnikov integral has isolated simple roots located at $\Omega\, \tau_0=(n-1/2)\pi$, for every $n$ integer (note that $K$ is also a function of $r_{un}$).
Within the Melnikov framework, this means that the system presents homoclinic intersections.

\begin{figure}
	\begin{center}
		\includegraphics[width=0.45\columnwidth]{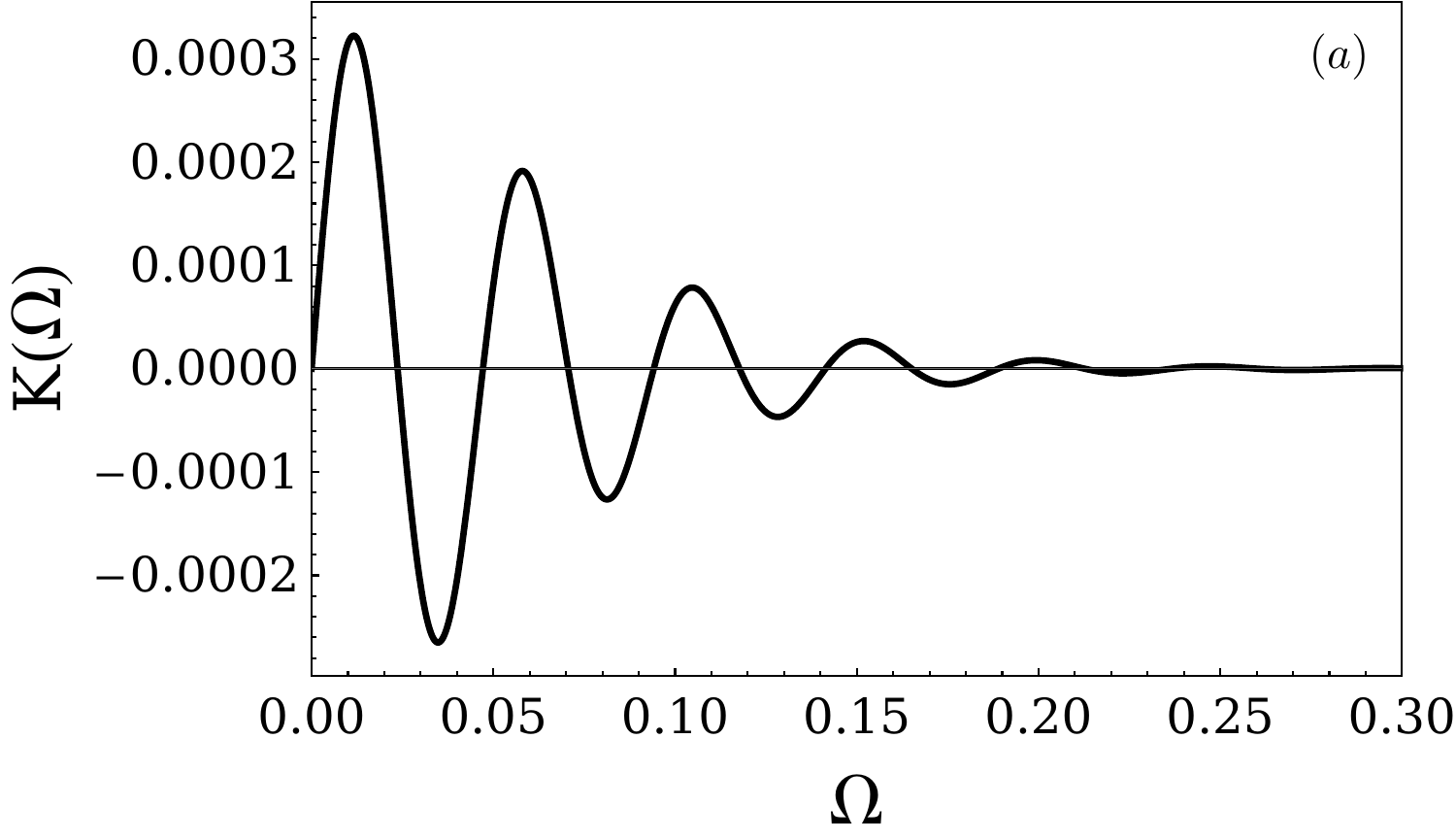} \qquad
		\includegraphics[width=0.45\columnwidth]{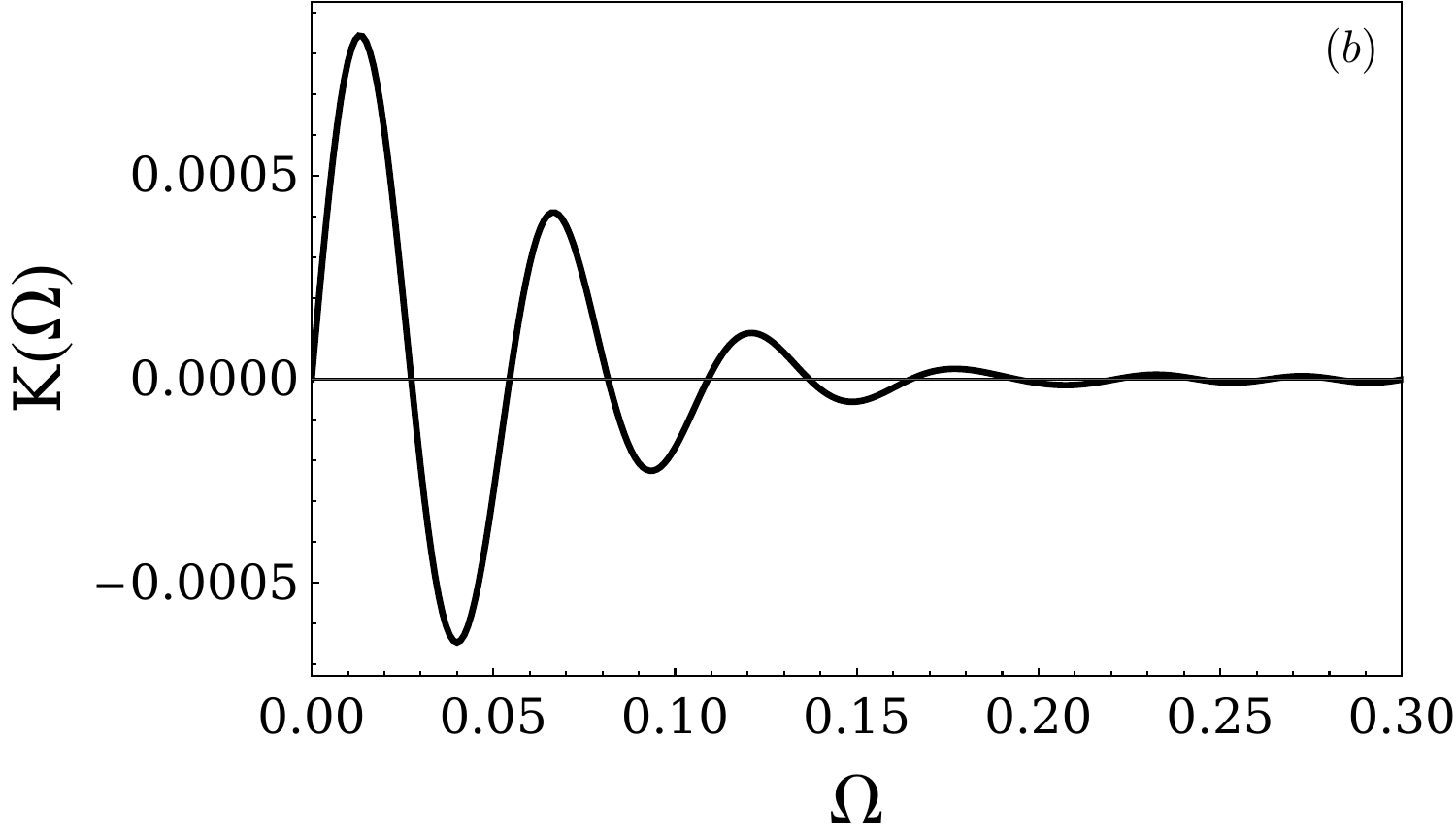}
		\caption{ Plot of $K(\Omega)$, with $M=1$, $Q=0.3$ and $r_{un}=4.5$ in the
		(a) Reissner-Nordstr\"om metric and the
		(b) Ay{\'o}n-Beato-Garc{\'i}a metric.
		}
		\label{fig:komega}
	\end{center}
\end{figure}

We note that these qualitative features, in particular the existence of a homoclinic tangle associated with the unstable fixed point of the stroboscopic Poincar\'e section, are typical for a generic choice of parameters of the system.

\subsection{Ay{\'o}n-Beato-Garc{\'i}a spacetime}
\label{secabg}

In the Ay{\'o}n-Beato-Garc{\'i}a spacetime the metric corresponds to (\ref{metric}) with \cite{ayonbeatoGarcia1998PRL}
\begin{equation}
\label{abgmetric}
 f(r)=1-\frac{2Mr^2}{(r^2+Q^2)^{3/2}}+\frac{Q^2r^2}{(r^2+Q^2)^{2}},
\end{equation}
where $M$ and $Q$ are the mass and charge of the black hole, respectively.
A charged black hole solution exists for $Q < Q_c \approx 0.634 M$, with two horizons located at $r=r_{\pm}$ \cite{ayonbeatoGarcia1998PRL}. In this case, as in the Reissner-Nordstr\"om black hole, homoclinic orbits associated with an unstable circular orbit of radius $r_{un}$ exist for $r_{boun}<r_{un}<r_{isco}$.  Figure~\ref{fig:orbitas}(b) shows the location of $r_{boun}/M$ (dashed line), $r_{isco}/M$ (dash-dotted line), together with $r_{+}/M$ (dotted line) for $0< Q/M < Q_c/M$.

Similarly to the previous case, the point-particle energy conservation equation for the homoclinic orbit may be cast in the following form:
\begin{equation}
 \label{eq:homoclinicaabg}
 \frac{dr}{d\tau}=\pm \sqrt{d^2-(r^2+w^2)\left(\frac{1}{r^2}+\frac{Q^2-2M\sqrt{Q^2+r^2}}{(Q^2+r^2)^2} \right)}
\end{equation}
where $d$ and $w$ are given by equations (\ref{eq:hk_a}) -- (\ref{eq:hk_b}).
Figure~\ref{fig:homoclinica}(b) shows the plot of $r(\tau)$ with $M=1$, $Q=0.3$, and $r_{un}=4.5$.

With the split of the homoclinic orbit due to the quadrupolar perturbation, we again apply the Melnikov approach, now with metric (\ref{abgmetric}), finding equations which differ from (\ref{eq:melnikovrn}) and (\ref{eq:integralrn}) solely by the function $k(r)$,  which is now given by equation (\ref{eq:k}).
In this case, the behavior of $K(\Omega)$ is illustrated in figure \ref{fig:komega}(b), with $M=1$, $Q=0.3$ and $r_{un}=4.5$. Again this shows the appearence of transverse homoclinic intersections due to the quadrupole perturbation.

As in Reissner-Nordstr\"om spacetime, the behavior of $K(\Omega)$ presented in figure \ref{fig:komega}(b) is typical for a generic choice of parameters of the system.

\section{Conclusion}
\label{sec:conclusion}

We analyzed the motion of spinless spherical test bodies in spherically symmetric spacetimes with metric given by equation~(\ref{metric}). We showed that the motion is in general not geodesic, except for the case of a cosmological vacuum. 
In particular, for a spherical test body with oscillating radius, the force term due to the body's quadrupole structure generally breaks the homoclinic orbits associated with unstable fixed points of test-particle motion, giving rise to homoclinic chaos. We presented examples of this phenomenon for the Reissner-Nordstr\"om and Ay\'on-Beato-Garc\'ia black hole spacetimes.

The procedure presented here, however, is far more general; embodied in our discussion is the derivation of the dynamics of generic spherical (spinless) test bodies moving in spacetimes with metrics of the form (\ref{metric}), for arbitrary choices of $f(r)$.
By analyzing the point-particle effective potential (\ref{eq:Veff}) for various naked-singularity spacetimes \cite{puglieseQuevedoRuffini2011PRD, vieiraMarekEtal2014PRD, stuchlikSchee2014CQG, stuchlikHledik2002AcPSl}, including the corresponding parameter region of Reissner-Nordstr\"om spacetime \cite{puglieseQuevedoRuffini2011PRD}, we note that the existence of unstable circular orbits with $e<1$ is a generic phenomenon for these no-horizon objects, at least for a given range of the metric parameters. This is also the case of the no-horizon parameter region of the Ay\'on-Beato-Garc\'ia spacetime \cite{garciaHackmannEta2015JMP}. A qualitative difference from the black-hole case is that in the absence of a horizon the centrifugal term in $V_{\rm eff}$ dominates near $r=0$, forming an inner centrifugal barrier outside the central singularity. The unstable orbits mentioned above are generally associated with homoclinic orbits in the $(r, p_r)$ phase space, a hint for the appearance of homoclinic chaos if the particle is substituted by an oscillating sphere.
We leave the discussion of these extended-body effects in naked-singularity spacetimes for future research.

\backmatter

\bmhead{Acknowledgments}
R.A.M. was partially supported by Conselho Nacional de Desenvolvimento Cient\'{i}fico e Tecnol\'{o}gico under grant Nos. 310403/2019-7 and 316780/2023-5.  F. F .R. was financed by the Coordena\c{c}\~{a}o de Aperfei\c{c}oamento de Pessoal de N{\'i}vel Superior - Brasil (CAPES) -~ Finance Code 001.

\noindent

\begin{appendices}

\section{Equations} \label{appendix}

The canonical equations of motion in Dixon's formalism and up to quadrupole order were given in (\ref{eq:sistdyn2}) in terms of the quadrupole moments $j_m(\tau)$ and $j_s(\tau)$, in addition to the functions $f_1(r,p_r)$, $f_2(r,p_r)$, $g_{1m}(r,p_r)$, $g_{1s}(r,p_r)$, $g_{2m}(r,p_r)$ and $g_{2s}(r,p_r)$. In this Appendix we provide the explicit forms of these functions:

\begin{eqnarray}
\label{eq:f1&g1&f2&g2m&g2s}
 f_1 &=&
\frac{r f(r)^{3/2} p_r}{\sqrt{r^2 p_t^2-f(r) \left(r^2 f(r) p_r^2+p_{\phi }^2\right)}},
\label{eq:f1} \\[3mm]
 f_2&=&
\frac{2 f(r)^2 p_{\phi }^2-r^3 f'(r) \left(f(r)^2 p_r^2+p_t^2\right)}{2 r^2 f(r)^{3/2} \sqrt{r^2 p_t^2-f(r) \left(r^2 f(r) p_r^2+p_{\phi }^2\right)}},
\label{eq:f2} \\[3mm]
 g_{1m} &=&
\frac{2 f(r)^3 p_r p_{\phi }^2 \left(r^2 f''(r)-2 f(r)+2\right)}{3 \left(r^2 p_t^2-f(r) \left(r^2 f(r) p_r^2+p_{\phi }^2\right)\right){}^2},
\label{eq:g1m}\\[3mm]
 g_{1s} &=&
g_{1m},
\label{eq:g1s}\\[3mm]
 g_{2m}&=&
\frac{D}{r}\! \Big[
r^2 f(r) p_t^2 p_{\phi }^2\! \left(r^3 f^{(3)}(r)+6 r^2 f''(r)+8\right)\!
+\!f(r)^4\! \left(8 r^2 p_r^2 p_{\phi }^2-r^5 p_r^4 \left(r \left(r f^{(3)}(r)
\right.\right.\right.\nonumber\\[2mm]
 &&\left.\left.\left.+2 f''(r)\right)-2 f'(r)\right)\right)-r^3 p_t^2 \left(r^3 p_t^2 \left(r f^{(3)}(r)+2 f''(r)\right)
+f'(r) \left(r^2 p_{\phi }^2 f''(r)
\right.\right.\nonumber\\[2mm]
 &&\left.\left.+2 \left(p_{\phi }-r p_t\right) \left(r p_t+p_{\phi }\right)\right)\right)+f(r)^3 p_{\phi }^2 \left(4 \left(p_{\phi }^2-2 r^2 p_r^2\right)-r^3 p_r^2 \left(r \left(r f^{(3)}(r)\right.\right.\right.\nonumber\\[2mm]
 &&\left.\left.\left.+6 f''(r)\right)-4 f'(r)\right)\right)-f(r)^2 \left(-2 r^7 f^{(3)}(r) p_r^2 p_t^2+2 r^2 f''(r) \left(p_{\phi }^4-2 r^4 p_r^2 p_t^2\right)
\right.\nonumber\\[2mm]
 &&\left.+r^3 p_r^2 f'(r) \left(r^2 p_{\phi }^2 f''(r)+4 r^2 p_t^2+2 p_{\phi }^2\right)
 +8 r^2 p_t^2 p_{\phi }^2+4 p_{\phi }^4\right) \Big],
  \label{eq:g2m}\\[3mm]
 g_{2s} &=&
D \Big[
2 r^3 f(r)^4 p_r^4 \left(r^2 f''(r)+2\right)+r^3 f(r)^3 p_r^2 \left(p_{\phi }^2 \left(r f^{(3)}(r)+2 f''(r)\right)+8 p_t^2\right) \nonumber\\[2mm]
 &&+r^2 p_t^2 \left(r^2 f''(r)+2\right) \left(2 r p_t^2-p_{\phi }^2 f'(r)\right)-r^2 f(r) p_t^2 \left(p_{\phi }^2 \left(r \left(r f^{(3)}(r)+2 f''(r)\right)
\right.\right.\nonumber\\[2mm]
 &&\left.\left.-4 f'(r)\right)+4 r p_t^2\right)-f(r)^2 \left(r \left(-r f^{(3)}(r) p_{\phi }^4-2 f''(r) \left(p_{\phi }^4-2 r^4 p_r^2 p_t^2\right)
\right.\right.\nonumber\\[2mm]
 &&\left.\left.+8 r^2 p_r^2 p_t^2\right)+p_{\phi }^2 f'(r) \left(r^4 p_r^2 f''(r)+2 r^2 p_r^2+2 p_{\phi }^2\right)\right)-4 r^3 f(r)^5 p_r^4 \Big],
 \label{eq:g2s}
\end{eqnarray}
where $D= \left[3 \Big( r^3 p_t^2-r f(r) \left(r^2 f(r) p_r^2+p_{\phi }^2\right)\Big)^2 \right]^{-1}$.
\vskip\baselineskip 
As explained in the main text, the unperturbed homoclinic orbit is associated with an unstable circular orbit of radius $r_{un}$. In the case of Ay{\'o}n-Beato-Garc{\'i}a spacetime, the homoclinic orbit is given in (\ref{eq:homoclinicaabg}) in terms of the parameters $d$ and $w$ shown below, which depend on $r_{un}$:
\noindent
\hspace*{-0.9cm}
\begin{minipage}{1.065\textwidth}
 \begin{eqnarray}
  \label{eq:hk}
 w&=&
   \frac{r_{un}^2\sqrt{Q^2(Q^2-r_{un}^2)\sqrt{Q^2+r_{un}^2}+M(r_{un}^4-2Q^4-Q^2r_{un}^2)}}{\sqrt{\sqrt{Q^2+r_{un}^2}(r_{un}^6+Q^6+3Q^4r_{un}^2+5Q^2r_{un}^4)-3M(Q^2+r_{un}^2)r_{un}^4}},\label{eq:hk_a} \\[3mm]
 d&=&\!
  \frac{\sqrt{\left(2Mr_{un}^2\sqrt{Q^2+r_{un}^2}-(r_{un}^4+Q^4+3Q^2r_{un}^2)\right)^2}}{\!\sqrt{\!\sqrt{Q^2\!+r_{un}^2}\!\left(\!\sqrt{Q^2\!+\!r_{un}^2}(r_{un}^6\!+\!Q^6+\!3Q^4r_{un}^2\!+\!5Q^2r_{un}^4)\!-\!3M(Q^2\!+\!r_{un}^2)r_{un}^4\right)}}.\label{eq:hk_b}
\end{eqnarray}
\end{minipage}
\vskip\baselineskip

Following the procedure described in the main text, the $k(r)$ functions in the Melnikov integral, (\ref{eq:integralrn}), can be determined, resulting in
 \begin{equation}
   k(r)=\frac{8Q^2\left[Q^2(3r_{un}^2-2r^2)-r_{un}(r^2r_{un}+3M(r_{un}^2-r^2))\right]}{3r^7(2Q^2+r_{un}(r_{un}-3M))},   \label{eq:kappa}
 \end{equation}
for the  Reissner-Nordstr{\"o}m spacetime, and
 \begin{eqnarray}
 k(r)&=&\frac{2Q^2r}{3(Q^2+r^2)^{11/2}}\Big[15M(4Q^4-3r^4+Q^2r^2)-4\sqrt{Q^2+r^2}(10Q^4+r^4-13Q^2r^2) \nonumber \\
   &&-3w^2(4\sqrt{Q^2+r^2}(r^2-7Q^2)+35M(Q^2+r^2))\Big], 
   \label{eq:k}
 \end{eqnarray}
for the Ay{\'o}n-Beato-Garc{\'i}a spacetime.




\end{appendices}




\end{document}